\documentclass{ws-jin}
\usepackage{url,wsnatbib}
\usepackage{graphicx}
\usepackage{multirow,bigstrut}
\usepackage{xcolor}

\begin{document}


\catchline{}{}{}{}{}

\title{fMRI Based Cerebral Instantaneous Parameters for Automatic Alzheimer's, Mild Cognitive Impairment and Healthy Subject Classification}

\author{Esmaeil Seraj*}

\address{School of Electrical and Computer Engineering, Georgia Institute of Technology\\
Atlanta, GA, United States\\
\email{eseraj3@gatech.edu}
}

\author{Mehran Yazdi*}

\address{Department of Communications and Electronics Engineering, School of Electrical and Computer Engineering, Shiraz University\\	Shiraz, Iran\\
		\email{yazdi@shirazu.ac.ir}
}

\author{Nastaran Shahparian}

\address{Department of Computer Science \& Engineering and Information Technology, School of Electrical and Computer Engineering, Shiraz University\\
	Shiraz, Iran\\
\email{nastaran.shahparian@gmail.com}
}

\author{*\textit{Corresponding Authors:} School of Electrical and Computer Engineering, Mollasadra Street, Shiraz University, Shiraz, Iran}

\maketitle

\begin{history}
\received{15 April 2019}
\revised{: \textcolor{blue}{Manuscript Submitted to be Considered By Journal of Integrative Neuroscience}}
\end{history}

\begin{abstract}
Automatic identification and categorization of Alzheimer's patients and the ability to distinguish between different levels of this disease can be very helpful to the research community in this field, since other non-automatic approaches are very time-consuming and are highly dependent on experts' experience. Herein, we propose the utility of cerebral instantaneous phase and envelope information in order to discriminate between Alzheimer's patients, MCI subjects and healthy normal individuals from functional magnetic resonance imaging (fMRI) data. To this end, after performing the region-of-interest (ROI) analysis on fMRI data, different features covering power, entropy and coherency aspects of data are derived from instantaneous phase and envelope sequences of ROI signals. Various sets of features are calculated and fed to a sequential forward floating feature selection (SFFFS) to choose the most discriminative and informative sets of features. A Student's t-test has been used to select the most relevant features from chosen sets. Finally, a K-NN classifier is used to distinguish between classes in a three-class categorization problem. The reported performance in overall accuracy using fMRI data of 111 combined subjects, is 80.1\% with 80.0\% Sensitivity to both Alzheimer's and Normal categories distinction and is comparable to the state-of-the-art approaches recently proposed in this regard. The significance of obtained results was statistically confirmed by evaluating through standard classification performance indicators. The obtained results illustrate that introduced analytic phase and envelope feature indexes derived from the ROI signals are significantly discriminative in distinguishing between Alzheimer's patients and Normal healthy subject.
\end{abstract}


\section{Introduction}
\label{sec:introduction}
Alzheimer is a chronic neurodegenerative disease that causes one's mental abilities such as memory and cognitive skills gradually decline, over the years. This occurs because of the reduction of healthy neurons involved in cognitive skills and as a result the atrophy of the brain. People generally are divided into three cases regarding to Alzheimer's disease, namely healthy subjects, Mild Cognitive Impairment (MCI) patients, and Alzheimer patients. MCI is a middle stage and a patient in this stage is at an increased risk of developing Alzheimer's or another dementia \cite{alzheimer20152015}.

Functional magnetic resonance imaging (fMRI) is a functional neuroimaging procedure using MRI technology which measures brain activity by detecting changes associated with blood oxygenated level dependent (BOLD) signal. There are two main approaches in studying fMRI: task-related fMRI, and resting state fMRI in which patients lying on the scanner with open eyes.
fMRI scans should be considered as a function of time, i.e. treat them as a time series(each time point representing one scan). This is because the BOLD signal will tend to be correlated across successive scans, meaning that they can no longer be treated as independent samples. The main reason for this correlation is the fast acquisition time (TR) for fMRI relative to the duration of the BOLD response.

Studies showed that neurodegenerative diseases such as Parkinson, Multiple Sclerosis (MS) and Alzheimer's can be observed the most significant on Default Mode Network (DMN) of the brain. By applying stimulations, energy consumption of brain increases approximately 5\% \cite{prvulovic2011functional}, and hereby, the rs-fMRI has been increasingly used in recent years as a noninvasive method in neuroimaging.

In general, Alzheimer identification methods are divided into two main groups: model based methods, and model-free methods. In the former, the goal is calculation of functional connectivity between anatomical or functional regions, For instance, Koch \cite{koch2012diagnostic} and Challis \cite{challis2015gaussian} applied seed based method in which time series correlations between a specific region and others are calculated. Although it is an easy method to be applied, finding the primary region is critical in reaching proper \cite{van2010exploring}.

In the model-free methods we try to estimate time series of voxels based on a reduced set of basis. Among these methods PCA and ICA are the most popular ones. In PCA, the goal is finding the correlated regions of voxels \cite{zhang2015functional}, while in ICA the goal is finding the independent sources. Since in PCA an optimal result occurs when the data have the normal probability density function (pdf), where fMRI data have not, the best usage of PCA is limited in filtering the noise in fMRI data \cite{ashby2011statistical}. On the other hand, the most important challenge in applying these two methods is finding the proper number of components \cite{binnewijzend2012resting}, for instance ICA finds the spatially independent component, blindly and at the end, hundreds of components might be found while only a couple of them are related to the study. 
Graph analysis is another model-free method in analyzing fMRI data. In this method, nodes are defined by anatomical or functional atlases and the weight of the edges are calculated by considering different criteria. A critical challenge here is defining the nodes and calculation of the weights of the edges, since different algorithms in calculating these two, can lead to different results \cite{bahrami2015assortativity, wang2010graph}.

Another method of analyzing is clustering method, during which the data are divided into subgroups having the most inter-group similarity and least intra-group similarity. Various kinds of clustering have been applied on fMRI data, for instance, Chen \cite{chen2012clustering} applied a hierarchical clustering method to define the difference in functional connectivity between MCI patients and healthy subjects. Their report showed that the distribution of clusters and their functionally disconnected regions are resembled to the altered memory network regions identified in task of fMRI studies. Clustering is easy to apply but it is time consuming for large databases such as fMRI. Moreover, defining the number of centers, determining a suitable distance criterion and performing an optimization strategy are so critical in this method. 

As recent endeavors to leverage fMRI data to investigate Alzheimer's and understanding the underlaying neuro-dynamics Zhu and Wang \cite{zhu2018exploring} proposed a supervised structure learning method to explore latent structures of resting state fMRI data belonging to different groups. The results reported a 'TREE' structure identified as the potential path for the progression of the Alzheimer's disease. In other studies such as \cite{golbabaei2016classification, khazaee2014automatic, golbabaei2016measures, lee2012resting} different machine learning and dictionary learning approaches are introduced and discussed to discriminate AD and MCI subjects. Nevertheless, these studies have used different network construction methods and are time-consuming and often require training on large datasets. In particular, many of these studies have significant differences in network construction methods (i.e., weighted versus binary and different density thresholds). It has been discussed before in \cite{reijneveld2007application} and \cite{fornito2010network, boostani2017comparative} that these differences are highly likely to affect the achieved results. Recently Wang et al \cite{wang2018classification} proposed an approach to discriminate between Alzheimer's diseases (AD) patients and MCI subjects under size limited fMRI data. The proposed method employs ROI analysis to derive correlation coefficient between various ROIs and then uses a regularized linear discriminant analysis (LDA) alongside with AdaBoost classifier to classify AD versus MCI subjects. We benchmark this study against our proposed procedure and present the discussion in the last section of this paper. In general, our approach leverages feature vectors and classification procedures that are more readily available and yet achieves comparable significant results \cite{sameni:hal-01355465, karimzadeh2015presenting}.

In order to perform an efficient and at the same time a simple analysis on the fMRI data, recently, the region-of-interest (ROI) analysis has been widely used \cite{poldrack2007region}. ROI analysis is a common approach to analyze the fMRI data in which signals from specified regions of interest (ROI's) are extracted.
ROIs can be extracted either in terms of structural or functional features. Structural ROIs are mostly defined based on macro anatomy, such as gyrus anatomy; whereas functional ROIs are generally based on analysis of data from the same individual. One common approach is to use a separate localizer scan to identify voxels that show a particular response in a particular anatomical region and then these voxels are explored to examine their response to some other manipulation.
When using single-subject atlases such as the AAL atlas or Talairach atlas in order to extract ROIs, one should be cautious about the inability of spatial normalization to perfectly match brains across individuals. Accordingly, the best practice is to use ROIs based on probabilistic atlases of macroscopic anatomy or probabilistic atlases which are available as part of the SPM Anatomy Toolbox or FSL \cite{poldrack2007region} In ROI analysis, by considering fMRI data as time series, the summation of time series of all voxels in specified anatomical or functional regions provide the ability of statistical analysis in signal processing terms.


In this study phase and envelope (amplitude) of ROI signals are used to present efficient, comprehensive and discriminative feature sets for the application of identifying Alzheimer's patients. For this purpose, instantaneous phase and envelope of ROI signals are estimated through analytic form representation for the sequences relating to each brain area. For instantaneous parameters, i.e. phase, frequency and envelope, estimating a recently proposed method named Transfer Function Perturbation (TFP) is used \cite{seraj2017robust}. TFP improves the quality of estimated instantaneous parameters by employing a statistical Monte Carlo based approach and removing the side-effects of previous conventional phase estimation methods \cite{sameni:hal-01355465}. After calculating the phase and envelope for brain areas in ROI signals, three types of features are introduced and estimated. Power, entropy and coherency are the main categories of estimated feature sets for both phase and envelope. Afterwards, a Sequential Forward Floating Feature Selection (SFFFS) algorithm is used to help choosing the most discriminative and informative sets of features among the introduced sets \cite{pudil1994floating}. Accordingly, Student's t-test is used in order to select the most relevant features. Eventually, K-Nearest Neighbors (KNN) classifier is employed to discriminate between three classes of (1) Alzheimer's patients, (2) MCI subjects and (3) Healthy normal individuals.

The rest of this study is structured as follows: within next section, first utilized dataset are introduced. Afterwards, the presented approach for calculating different feature sets is detailed and each step is elaborated separately. Finally, the results are represented and discussed in last two sections.

\section{Methodology}
\label{sec:method}
\subsection{Dataset}
\label{subsec:dataset}
Rs-fMRI and high resolution T1-weighted MRI images obtained from the Alzheimer's disease neuroimaging initiative (ADNI) database \cite{jack2008alzheimer} Data from 111 subjects gained, in which 43 data belong to healthy normal subjects, 36 to MCI patients and 32 to Alzheimer patients. Table 1 shows the detailed information of each group. For each subject, 140 gradient echo planar imaging (EPI) volumes were acquired by using 3T Phillips Scanner. The parameters of the scanner are TR = 3s, TE = 30ms, matrix size = $ 64\times64 $, slice thickness = 3mm, and number of slices = 48.

 \begin{table}[tbh]
 	\centering
 	\caption{Subject Cohort}	
 	\begin{tabular}{c|c|c|c}
 	& Normal & MCI & Alzheimer \\
 	\hline
 	\hline
 	Number of subjects &	43 &	36 &	32 \\
 	Male/Female	& 17/26 &	14/22 &	15/17 \\
 	Mean Age &	75.30 &	72.75	& 72.34 \\
 	Standard deviation Age &	6.37 &	6.35	& 7.12 \\
 	Mean Education &	16.27 &	15.25 &	15.75 \\
 	Standard deviation Education &	2.1 &	2.54 &	2.75 \\

 	\end{tabular}
 	\label{tab:SubjectCohort}
 \end{table}
 
\subsection{Preprocessing: Extracting ROI Signals}
\label{subsec:PreprocessingExtractingROISignals}
\begin{figure}[tb]
\centering
\includegraphics[trim=0 0in 0in 0in,clip,width=\columnwidth]{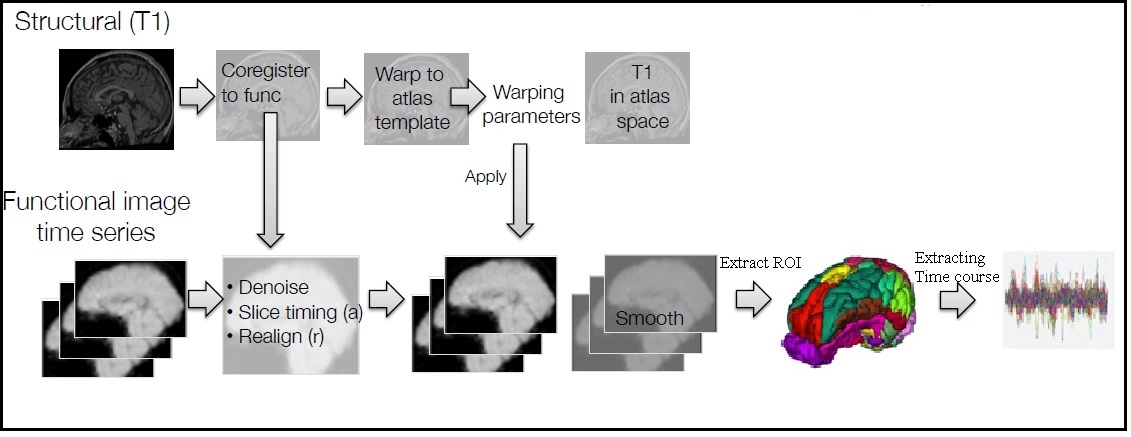}
\caption{Employed procedure for ROI signal extraction from raw fMRI data.}
\label{fig:PreprocessingROIExtraction}
\end{figure}

All processes have been carried out by using FSL (fMRIB's Software LibraryUK), REST toolbox (Developed by Zhang \textit{et al}. at Laboratory of Cognitive Neuroscience and Learning, Beijing Normal University, China), and MATLAB programming environment. Here we bring the details of each step in the preprocessing procedure. For a better understanding, a complete elaboration of the procedure is also represented in Fig.~\ref{fig:PreprocessingROIExtraction}.

The applied preprocessing steps can be summarized as follow:

\begin{enumerate}
\item Applying head movement correction.
\item Applying slice timing correction.
\item Applying a spatial filter by using an 3D Gaussian kernel with 4mm3 FWHM in order to increase accuracy of registered functional images to standard space and to achieve better signal to noise ratio (SNR)
\item Applying high pass filter with 100s cut-off frequency  to remove low level noise
\item Registering functional images to T1-weighted images and then registering to MNI152 space using the transformations calculated on corresponding anatomical images.
\item Applying a band pass filter (0.01HZ - 0.1HZ), since resting state BOLD signal, which arises from neuronal activity, is located in this frequency band.
\item Filtering output data in the previous step by regressing out movement vectors as suggested by Friston~\textit{et al} \cite{friston1996movement}.
\item Filtering unwanted signals such as physiological noise which can be conducted by  principle component analysis as it mentioned by Behzadi~\textit{et al} \cite{behzadi2007component}.
\item Filtering the linear trend of gray matter time series which occurred due to the heat of scanner.
\item Obtaining time series of 112 anatomical regions for each subject based on Harvard-oxford atlas in “Extract ROI time courses” tab in REST software.
\end{enumerate}

Harvard-oxford is a probabilistic atlas covering 48 cortical and 21 sub-cortical structural areas, derived from structural data and segmentations provided by the Harvard Center for Morphometric Analysis. In this atlas, T1-weighted images of 21 healthy male and 16 healthy female subjects (ages 18-50) were individually segmented using semi-automated tools. The T1-weighted images were affine-registered to MNI152 space and the transforms are then applied to the individual labels. Finally, they were combined across subjects to form population probability maps for each label~\cite{karimzadeh2017sleep}.

The summation of time series of these regions is reformed into a vector of $ 112\times1 $ for each subject and then by putting together these vectors for each group, a final matrix of $ 112\times16 $ is obtained and finally feature extraction method is applied on these matrices.

\subsection{Instantaneous Parameters Estimation}
\label{subsec:InstantaneousParametersEstimation}
The conventional approach for instantaneous phase, frequency and envelope estimation, i.e. narrowband frequency filtering followed by analytic or complex representation, is prone to highly affect the estimates and yield ambiguous values, especially during low SNRs of background activity \cite{seraj2017robust}, \cite{sameni:hal-01355465}. The background activity here is referred to the undesired components of the frequency-specific instantaneous measures.
For obtaining an accurate and unambiguous estimation of instantaneous phase (IP) and instantaneous envelope (IE), herein, we use a recently developed method in \cite{seraj2017robust} and \cite{sameni:hal-01355465}. The method which is referred to as \textit{Transfer Function Perturbation} (TFP) is a statistical Monte Carlo based estimation scheme in which infinitesimal perturbations or dithers are added to the utilized filter or input signal in order to generate estimation ensembles \cite{seraj2017robust}. The applied perturbations or dithers in TFP are such that they are physiologically irrelevant and the filter's specifications do not change significantly according to the estimation standards \cite{sameni:hal-01355465}. The filtering process in TFP is performed in a forward-backward zero-phase approach in order to prevent any phase distortion. Eventually, the final IP and IE estimates are calculated through ensemble averaging over all dithered and perturbed ensembles. The rationale behind the TFP is beyond the scope of the current study and one can find a detailed description in \cite{sameni:hal-01355465} and \cite{seraj2017robust}. To  date, TFP has been successfully used in a variety of applications such as BCI \cite{seraj2017robust, seraj2017improved}, brain connectivity and synchronization \cite{sameni:hal-01355465, seraj2017investigation} and sleep stage classification \cite{karimzadeh2018distributed}.
In this study, for both IP/IE estimation and also deriving relevant phase and envelope features, we use the \textit{cerebral signal phase analysis toolbox} provided by the authors of \cite{seraj2017robust, sameni:hal-01355465, seraj2017investigation} which is introduced in \cite{Seraj2016} and is available online at \cite{sameniopen}. Accordingly, the analytic representation for sequences relating to each brain area in ROI signals extracted from fMRI data is calculated as follows:
 \begin{equation}
 Z_i(t)=x_i(t)+j\mathcal{H}\{x_i(t)\}
 \label{eq:AnalyticForm}
 \end{equation}
 
Where $ x_i(t) $ is the sequence in $ i $-th brain area of extracted ROI signal and H{.} represents the Hilbert Transform. Using the represented analytic form the instantaneous phase ($ {IP}_i $) and envelope ($ {IE}_i $) for each brain area ($ i $) are estimated as follows:

 \begin{equation}
{IP}_i(t)=\arg\{Z_i(t)\}=\arctan\left(\frac{\mathcal{H}\{x_i(t)\}}{x_i(t)}\right)
 \label{eq:AnalyticPhase}
 \end{equation}
  \begin{equation}
{IE}_i(t)=\left|Z_i(t)\right|=\sqrt{x_i(t)^2+\mathcal{H}\{x_i(t)\}^2}
  \label{eq:AnalyticEnvelope}
  \end{equation}
  
\begin{figure}[tb]
 \centering
 \includegraphics[trim=0.05in 0in .75in 0.2in,clip,width=\columnwidth]{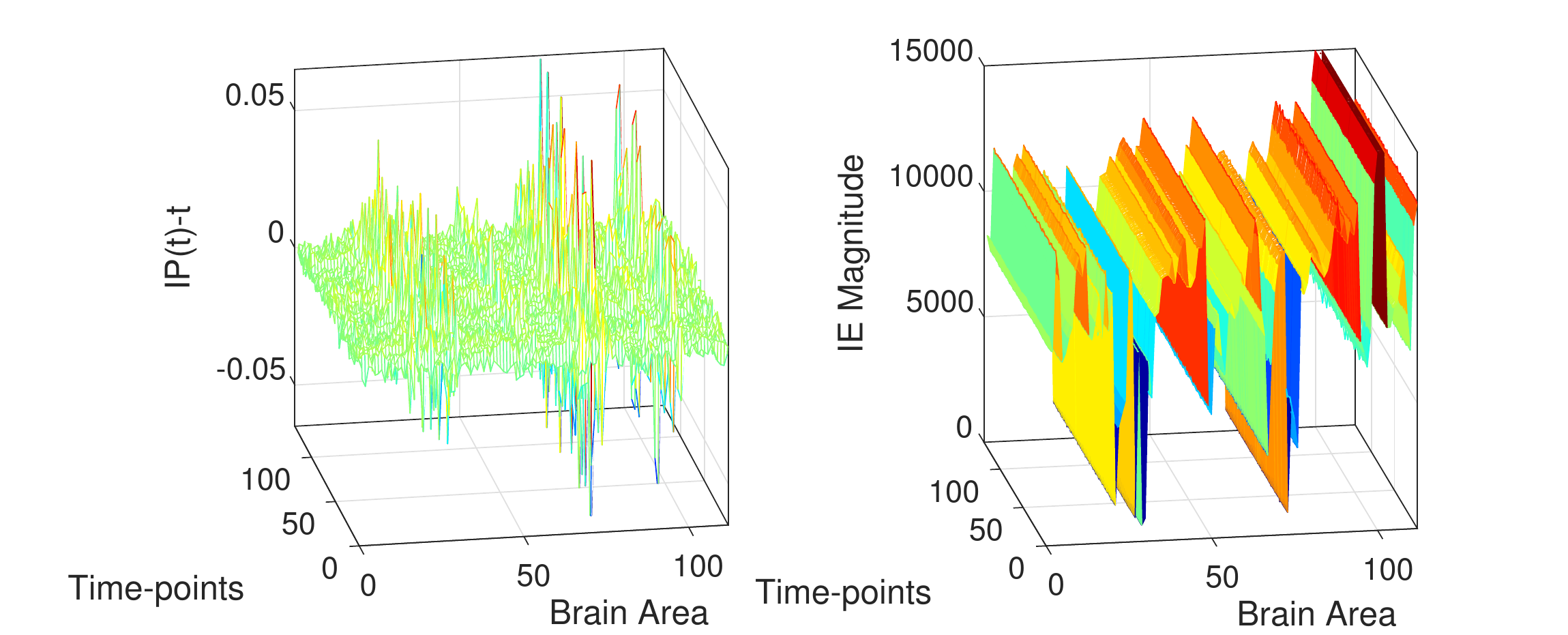}
 \caption{Estimated IP and IE for each brain area over time-points calculated for the ROI signal extracted from fMRI data of an Alzheimer's patient}
 \label{fig:IP_TE}
 \end{figure}

 Due to the usage of $ \arctan\left(.\right) $ function, the calculated phase signal might be wrapped in points where the values cross $ \pm\pi $. Accordingly, an unwrapping step is required after estimating the IP as a post-processing level. Fig.~\ref{fig:IP_TE} shows the estimated IP and IE for each brain area over time-points calculated for the ROI signal extracted from fMRI data of a subject in employed dataset.
 
 \subsection{Feature Estimation: Introducing Feature Indexes}
 \label{subsec:FeatureEstimation}
 The estimated phase and envelope measures for ROI signals derived from fMRI data of each subject, i.e. Alzheimer, MCI and Normal, are then used to extract the feature sets. Three different categories of features as (1) power, (2) entropy and (3) coherency are calculated for IP and IE to cover almost all aspects of physiological data by using both local-scale (relating to one specific area of brain) and large-scale (between two distant areas within brain) features.
 
 \subsubsection{Power Feature Sets}
 \label{subsubsec:PowerFeatureSets}
 Energy of the calculated IPs and IEs over time-points for each brain area ($ i $) is used as a measure of power, indicating a local-scale feature set. This feature is used to capture the intensity of brain activity in separate areas. Accordingly, different amount of activity recorded in each cerebral region could potentially be discriminative between Alzheimer's, MCI's and Normal subject's data. The Energy of ROI signals for each brain area ($ i $) over a period of T time-points can be calculated as follows:
 
 \begin{equation}
{IPPow}_i=\sum_{t=1}^{T}|{IP}_i(t)|^2
\label{eq:IPPow}
 \end{equation}
  \begin{equation}
 {IEPow}_i=\sum_{t=1}^{T}|{IE}_i(t)|^2
 \label{eq:IEPow}
  \end{equation}
  
  The calculated energy values are stored in vectors of size N which represents the number of brain areas. Accordingly, for each subject, two vectors of length 112 (N = 112) are computed as IP and IE power feature sets.
  
  \subsubsection{Entropy Feature Sets}
  \label{subsubsec:EntropyFeatureSets}
  Entropy indexes are directly related to the amount of information embedded in a signal. Herein, for capturing irregularity and significance of variations of the brain activity within different regions, \textit{Shannon Entropy} is used as another local-scale feature. Although variance and entropy indexes both reveal the information regarding variations and temporal irregularity of the patterns in a signal, the variance is sensitive to the amplitude values \cite{sabeti2009entropy}. Accordingly, using the estimated IP and IE images as illustrated in Fig. 3, Shannon Entropy can be calculated for separate brain areas as follows:
  
  \begin{equation}
{IPEnt}_i=-\sum_{k}p_k\log_bp_k
  \end{equation}
  \begin{equation}
  {IEEnt}_i=-\sum_{k}l_k\log_bl_k
    \end{equation}
    
where $ k $ is the range of all discrete amplitude values of the signals. Also,$ p_k $ and $ l_k $ are the probability of the$  IP_i(t)  $and $ IE_i(t) $signals having the $ k $-th magnitude, respectively. Histogram analysis is a proper technique to calculate the probabilities and the entropy in case that the number of samples in different discretized magnitudes are sufficient. Moreover, the ranges of IP and IE sequences are not equal, where consequently, the width of bins $ p_k $ and $ l_k $ are different and varied from one feature to another.

Similar to the first feature set, i.e. power features, the calculated entropies are stored in vectors of size N=112 representing the number of brain areas. Accordingly, for each subject in each of three classes, two vectors, i.e. for IP and IE, are computed as the second sets of features.

\subsubsection{Coherency Feature Sets}
\label{subsubsec:CoherencyFeatureSets}
Two different but inherently similar coherency indexes, namely Phase Locking value (PLV) and Magnitude Squared Coherence (MSC), are proposed here in order to investigate the correlation and dependence between apart regions of brain and create large-scale feature sets. PLV and MSC feature sets are calculated for the IP and IE sequences extracted from ROI signals, respectively.

PLV is one of the most common measures used in phase analysis which describes how much the difference between phases of two signals is constant \cite{lachaux1999measuring}. For calculating PLV, after estimating the IP difference between two signals, the local stability of this IP difference have to be quantified. Accordingly, the stability of IP differences between brain regions ($ i $) and ($ j $) can be quantified as below \cite{lachaux1999measuring}.

\begin{equation}
PLV_{ij}=\left|\frac{1}{T}\sum_{t=1}^{T}e^{j[{IP}_j(t)-{IP}_i(t)]}\right|
\end{equation}

where T is the length of signals and the summation is taken over time-points ($ t $). 

The MSC is employed to investigate the between-region coherency for estimated envelope (IE). The conventional approach for measuring MSC is based on calculating the Power Spectral Densities (PSD) for two signals \cite{carter1973estimation}. Assuming $ IE_i  $and $ IE_j  $to be two randomly chosen instantaneous envelope signals from two distant brain areas (i) and (j), the MSC can be computed as below \cite{seraj2016cerebral, carter1973estimation}:

\begin{equation}
MSC_{ij}=\frac{|PSD_{ij}|^2}{PSD_{ii}PSD_{jj}}=\frac{\mathbb{E}\{{IE}_i{IE}_j^*\}}{\mathbb{E}\{|{IE}_i|^2\}\mathbb{E}\{|{IE}_j|^2|\}}
\end{equation}

Where $ \mathbb{E}\{.\} $ is the mathematical expectation and $ PSD_{ij} $ is the cross-spectrum between instantaneous envelope sequences estimated from ROI signals extracted for brain areas ($ i $) and ($ j $) \cite{seraj2016cerebral}. PLV and MSC are both widely used cerebral synchrony indexes and their values vary between 0 and 1. A PLV or MSC equal to 1 indicates highly coherent and synchronous signals and vice versa \cite{seraj2016cerebral, carter1973estimation}. 

\begin{figure}[tb]
 \centering
 \includegraphics[trim=0.4in 0in 0.7in 0.25in,clip,width=\columnwidth]{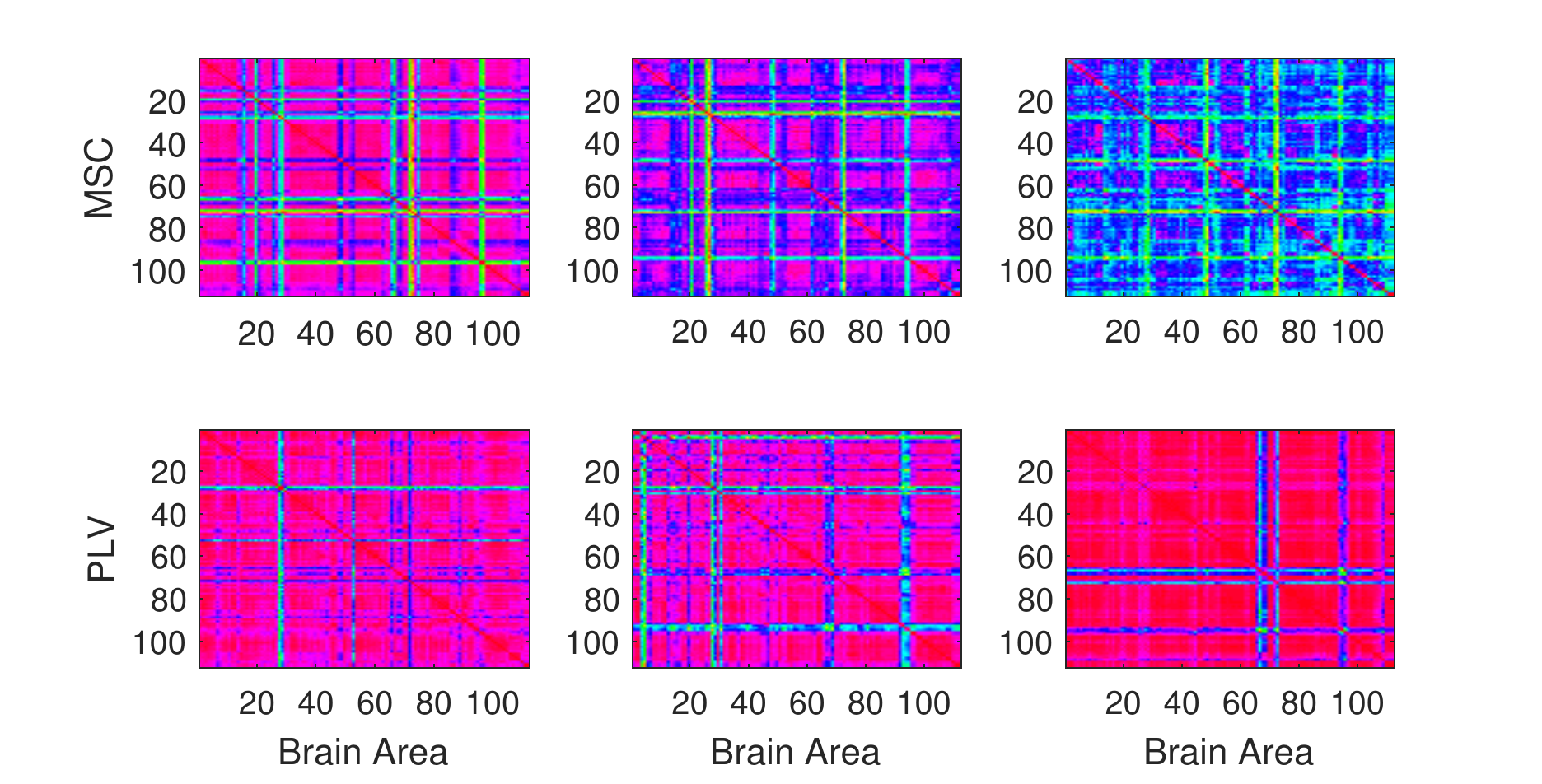}
 \caption{Sample pairwise PLV and MSC matrices calculated for IP and IE sequences, respectively. From left to right, illustrated matrices belong to Alzheimer's, MCI and Normal subjects from utilized dataset. The IPs and IEs are extracted from ROI signals for all (i.e. N = 112) brain areas.}
 \label{fig:MSC_PLV}
 \end{figure}

For both PLV and MSC, the coherency is inspected between all possible pairs of 112 brain areas, resulting in $ 112\times112 $ feature matrices for$  IP_i(t)  $ and $  IE_i(t)  $ respectively. Figure~\ref{fig:MSC_PLV} illustrates sample PLV and MSC matrices calculated for IP and IE sequences of a subject in utilized dataset. In this way, the third category of features is formed as a large-scale feature, covering the coherency and synchrony between the activities of different cerebral areas in fMRI data.

\subsection{Feature Selection and Classification}
\label{subsec:FeatureSelectionandClassification}
In this step, first, a \textit{Sequential Forward Floating Feature Selection} (SFFFS) algorithm is applied to identify the most informative and discriminative sets of features among all 6 feature sets (2 sets for IP and IE in each category). The feature vector for each class is formed by concatenating the features calculated from the corresponding IP and IE signals. In this step, each feature set is used solely is a classification process in similar settings and the weakest sets in accuracy are left out.

The chosen features and remaining sets from the former step are fed to Student's t-test in order to select the most relevant and discriminative features within feature-sets. Student's t-test assumes a normal distribution for the features of each class with equal but unknown variances and examines the null hypothesis of whether they have equal means \cite{duda1973pattern}. Accordingly, only features with p-values below a significance level equal to 0.05, indicating confidence level in the rejection of the null hypothesis, are included in the classification stage.

Eventually, the remaining features are gathered and concatenated for each of three classes and the corresponding labels are assigned. The Alzheimer's patients, MCI subjects and healthy Normals are labeled as 1, 2 and 3 respectively. The combination of all selected feature sets together is fed to a K-Nearest Neighbor (KNN) classifier with K=5 to perform the final discrimination between three classes. Although various other values of K have been tested, i.e. K=1,5,10 and 15, K=5 showed the best performance and was chosen for all levels of classification (i.e. in feature selection with SFFFS). For the classification, a total of 30 subjects' data (i.e. 10 of each class) have been chosen randomly for the test and the remaining 81 were used for training the classifier. Accordingly, the classifier is trained by the entire combined sets of features with a single label and then returns a single value, i.e. 1, 2 or 3, as the result of label testing with test data.

\section{Experimental Results}
\label{sec:results}
\begin{figure}[tb]
 \centering
 \includegraphics[trim=1.5in .85in 2in 2in,clip,width=\columnwidth]{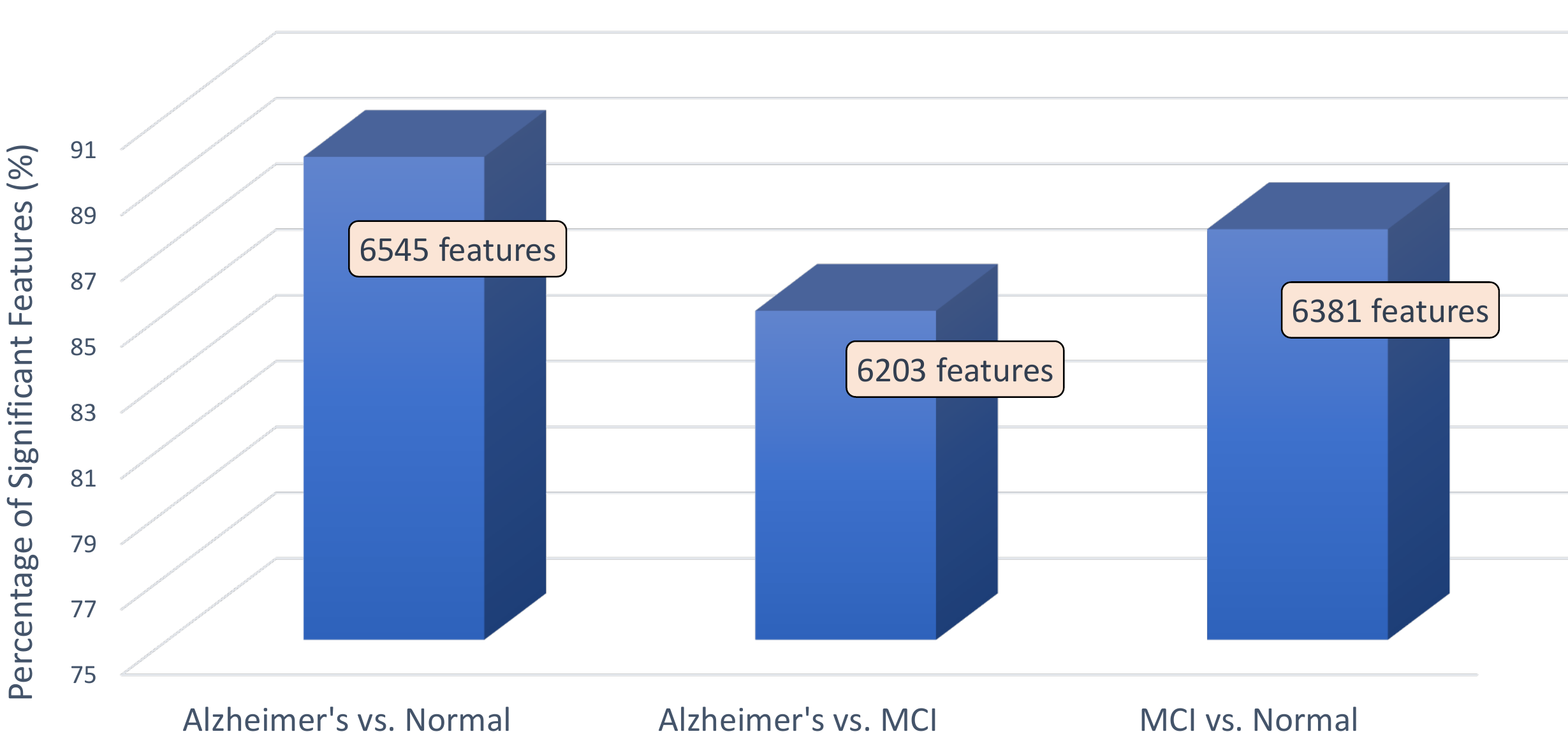}
 \caption{Amount of statistically significant features selected by Student's t-test to be involved in classification stage.}
 \label{fig:Significant_Features}
 \end{figure}

In this section, the results of classifying between Alzheimer's, MCIs and Normal subjects by the combination of all chosen features are evaluated through calculating four standard classification performance indicators, namely accuracy (AC), precision (PR), specificity (SP) and sensitivity (SE).First, it is noteworthy to review the results of feature selection through significance tests. The significance tests were performed for all three possible cases, i.e. Alzheimer's vs. MCI, Alzheimer's vs. Normal and MCI vs. Normal, and the results are elaborated in Fig.~\ref{fig:Significant_Features}. The confidence level for the rejection of the null hypothesis, as mentioned, was chosen equal to 5\% (p-value $ < $ 0.05). As it can be seen, approximately between 80\% to 90\% of the calculated features were confirmed as statistically significant and relevant.

 The selected features were involved in classification stage. By using a 5-NN classifier as described previously, we were able to correctly label 21 subjects out of selected 30 for the test, resulting in a 80.1\% overall accuracy (average accuracy of all classes). The confusion matrix of this evaluation is represented in Table~\ref{tab:CM}. 
 
  \begin{table*}[tbh]
  	\centering
  	\caption{Confusion matrix of the proposed automatic Alzheimer's detection method}	
  	\begin{tabular}{c|c|c||c||c|c}
  		& & \multicolumn{3}{c|}{Truth Data} & Precision \\
  		\cline{3-6}
  		
  		\parbox[t]{3mm}{\multirow{5}{*}{\rotatebox[origin=c]{90}{\small Classifier}}} 
  		&  & Alzheimer's & MCI& Normal & \\
  		\cline{1-6}
  		&	Alzheimer's & 8 &  3 & 1 & 66.7\% \\
  		\cline{2-6}
  		& MCI &  2 &  5 & 1 & 62.5\% \\
  		\cline{2-6}
  		& Normal &  0 &  2 & 8 & 80.0\% \\
  		\cline{1-6}
  		\multicolumn{2}{c|}{Recall (Sensitivity)} &  80.0\% &  50.0\% & 80.0\% & Overall Accuracy = 80.1\% \\
  	\end{tabular}
  	\label{tab:CM}
  \end{table*}
  
\begin{figure}[tb]
 \centering
 \includegraphics[trim=1in .9in 1.5in 0.8in,clip,width=\columnwidth]{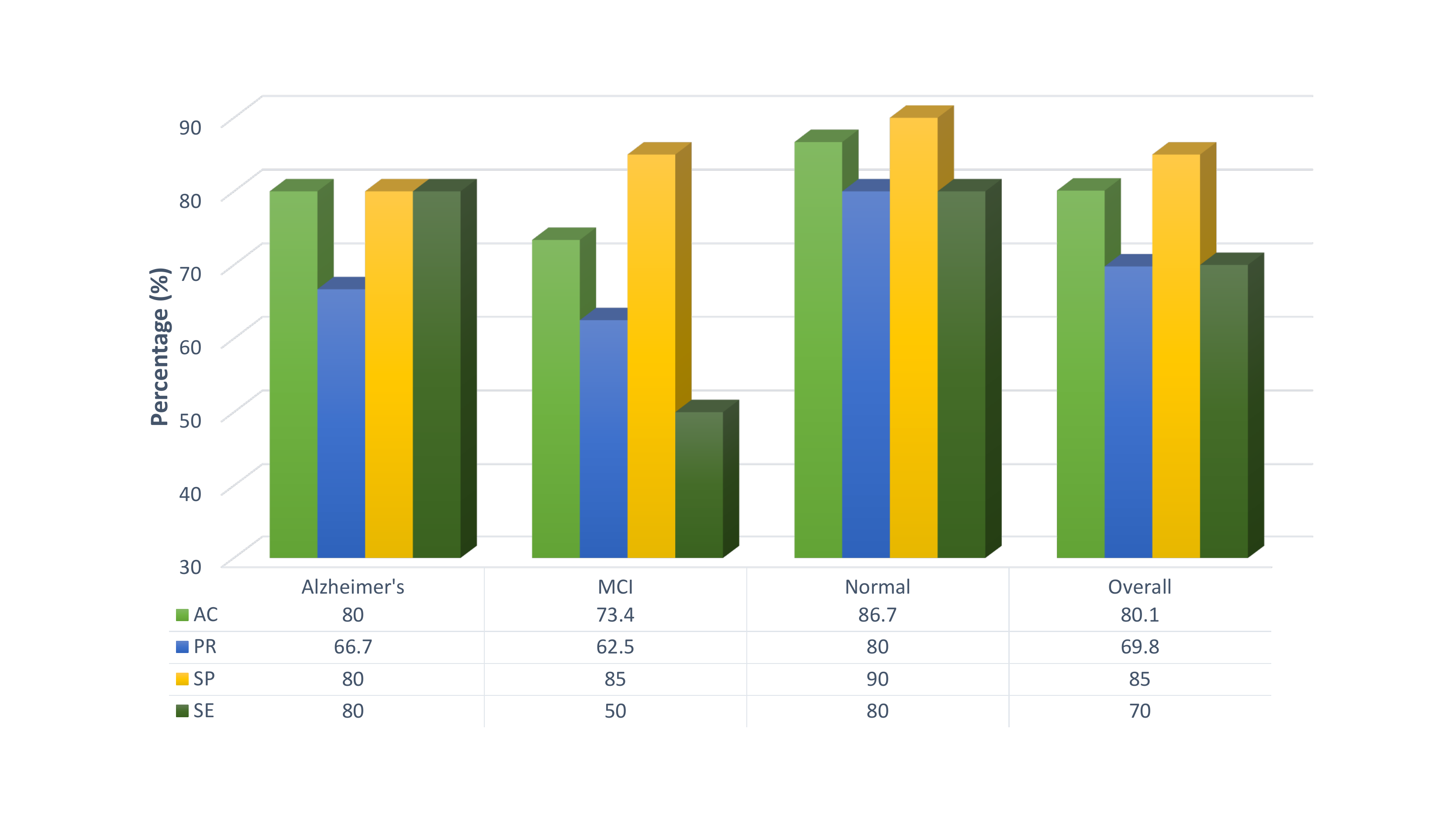}
 \caption{The obtained accuracy (AC), precision (PR), specificity (SP) and sensitivity (SE) for each of three class and the overall case.}
 \label{fig:Results_All}
 \end{figure}

Considering the result presented in Table~\ref{tab:CM}, the accuracy, precision, specificity and sensitivity of the proposed features are illustrated in Fig.~\ref{fig:Results_All} both for each class and in overall. The AC, PR, SP and SE are calculated as below:

\begin{equation}
AC=\frac{TP+TN}{TP+TN+FP+FN}
\label{eq:AC}
\end{equation}
\begin{equation}
PR=\frac{TP}{TP+FP}
\label{eq:PR}
\end{equation}
\begin{equation}
SP=\frac{TN}{TN+FP}
\label{eq:SP}
\end{equation}
\begin{equation}
SE=\frac{TP}{TP+FN}
\label{eq:SE}
\end{equation}

Accordingly, the overall accuracy, precision, specificity and sensitivity are then computed as the average AC, PR, SP and SE of all classes, respectively.

As depicted in Fig.~\ref{fig:Results_All}, the classes belonging to the Alzheimer's patients and healthy Normal subjects are showing great results, however, one might discuss that this is not the case for the class of MCI subjects. Table~\ref{tab:CM} states that 8 out of 9 occurred misclassifications are somehow related to MCI category. Moreover, 3 of MCI cases have been mistaken by Alzheimer's patients which shows close specifications between these two classes. As a consequent, although 8 out of 10 Alzheimer's subjects have been identified correctly, SP and PR are not relatively high for this class (as compared to the Normal category). Generally speaking, the proposed phase and envelope features are showing significant results; nevertheless, further strategies are required in order to improve the classification rate of MCI subjects.

\section{Discussion}
\label{sec:discussion}
Detection of Alzheimer disease in its early stage is significantly important for medicines to apply proper treatments. Therefore, new trends in this domain are toward using more efficient algorithms to distinguish normal subjects from Alzheimer patients. In this paper, we developed an algorithm which can efficiently classify subjects into normal and different stages of Alzheimer disease using fMRI data. To do that, we use for the first time new features which are phase and envelope sequences of ROI signals from fMRI data, and the selected ROI are based on Harvard-Oxford atlas which is a probabilistic brain atlas. We also selected most informative sets of these features using a sequential forward floating feature selection. Our observation showed that this new set of features can efficiently represent the characteristics of fMRI data and discriminate very well different stages of Alzheimer disease. As it shown in figure 6, separating MCI patients has the least accuracy and the algorithm mostly mistaken them by Alzheimer's and that is because of the variation of the brain in patients such as Aggregation of protein fragment beta-amyloid outside the neurons and also abnormally accumulation of protein tau (tau tangles) inside neurons, which is basically similar in MCI and Alzheimer patients rather than normal ones. In general the obtained results by applying the proposed algorithm on real fMRI data showed a good performance which can be a promising approach for clinical applications.

\section*{Acknowledgments}
This research has been supported by the Cognitive Sciences and Technologies Council of Iran (COGC), under the grant number 2250.

\section*{Conflict of interest statement}
The authors declare no competing interests.

\section*{Authors' Contributions}
ES proposed the algorithms and was responsible for data analysis and classification. NS was responsible for data preparation and preprocessing and also writing the Introduction and preprocessing sections. ES wrote the rest of the manuscript. MY was in charge of general research guidelines and idea development. All authors contributed to editorial changes in the manuscript. All authors read and approved the final manuscript.

\bibliographystyle{ws-jin}
\bibliography{References_AlzheimerPaper}

\end{document}